\documentclass[lineno]{jfm}

\usepackage{mathtools}
\usepackage{graphicx}
\usepackage{newtxtext}
\usepackage{newtxmath}
\usepackage{natbib}
\usepackage{hyperref}
\usepackage{ulem}
\hypersetup{
    colorlinks = true,
    urlcolor   = blue,
    citecolor  = black,
}
\newcommand{\RomanNumeralCaps}[1]
\linenumbers

\usepackage{scalerel}[2016-12-29]

\def\scalesum#1{\vcenter{\hbox{\scaleto[3ex]{\displaystyle\sum}{#1}}}}

\newcommand{\vel}{\boldsymbol{u}}
\newcommand{\pos}{\boldsymbol{x}}
\newcommand{\scl}{\boldsymbol{k}}

\newcommand{\varpos}{\boldsymbol{r}}

\newcommand{\fil}{\overline{\vel}^l}

\newcommand{\Tr}{\mbox{Tr}}

\title{Fluctuations around Turbulence Models}

\author{Flavio Tuteri\aff{1}
  \corresp{\email{flavio.tuteri@phys.ens.psl.eu}},
  A. Alexakis\aff{1}
 \and S. Chibbaro\aff{2}}

\affiliation{\aff{1}Laboratoire de Physique de l’Ecole normale supérieure, ENS, Université PSL, CNRS, Sorbonne Université, Université de Paris, F-75005 Paris, France
\aff{2}LISN}

\begin{document}

\makeatletter
\let\linenumbers\relax
\let\endlinenumbers\relax
\makeatother

\maketitle

% ----- ===== ----- ===== ----- ===== ----- ===== ----- ===== ----- ===== ----- ===== ----- ===== |

\begin{abstract}
Numerical simulations of turbulent flows at realistic Reynolds numbers generally rely on filtering out small scales from the Navier–Stokes equations and modeling their impact through the Reynolds stress tensor ${\tau}_{ij}$. Traditional models approximate ${\tau}_{ij}$ solely as a function of the filtered velocity gradient, leading to deterministic subgrid-scale closures.
%that fail to capture the intrinsic chaotic variability of the small-scale dynamics. 
However, small-scale fluctuations can locally exhibit instantaneous values whose deviation from the mean can have a significant influence on flow dynamics.
%that deviate substantially from the mean that can have a significant influence on flow dynamics. 
In this work, we investigate these effects by
employing direct numerical simulations combined with Gaussian filtering to quantify subgrid-scale effects and evaluating the local energy flux in both space and time. The mean performance of the canonical Clark model is assessed by conditioning the energy flux distributions on the invariants of the filtered velocity gradient tensor, $Q$ and $R$. The Clark model captures to a good degree the mean energy flux.
%which describe the topology of the filtered velocity field . 
%To move beyond mean analysis,
However, the fluctuations around these mean values for given ($Q,R$) are of the order of the mean displaying fat tailed distributions.
To become more precise, we examine the joint distributions of true energy flux and the predictions from both the Clark and the Smagorinsky models.
%the  by analyzing conditional PDFs at fixed model outputs.
This approach mirrors the strategy adopted in early stochastic subgrid-scale models.
%We further extend this analysis to the eddy-viscosity-based Smagorinsky model. 
Clear non-Gaussian characteristics emerge from the obtained distributions, particularly through the appearance of heavy tails. The mean, the variance, the skewness and flatness of these distributions are quantified.
Our results emphasize that fluctuations are an integral component of the small-scale feedback onto large-scale dynamics and should be incorporated into subgrid-scale modeling through an appropriate stochastic framework.
%\ALEX{Furthermore, they put a strong constrain on future stochastic models.}
\end{abstract}


% {\bf MSC Codes }  {\it(Optional)} Please enter your MSC Codes here

% ----- ===== ----- ===== ----- ===== ----- ===== ----- ===== ----- ===== ----- ===== ----- ===== |

\section{Introduction}
\label{sec:introduction}
Turbulent flows are the fundamental basis of many engineering applications \citep{pope2001turbulent}, geophysics \citep{thorpe2007introduction,gill2016atmosphere}, and astrophysics \citep{biskamp2003magnetohydrodynamic} among others.
Despite the considerable improvement in computational power, performing direct numerical simulations (DNS) that resolve all active scales in realistic turbulent fluids remains out of reach for most industrial and natural flows \citep{pope2001turbulent,moin1998direct}. 
That is related to the fact that the number of degrees of freedom grows fast %\sout{exponentially} 
with the size of the system. To circumvent this prohibitive cost, large-eddy simulations (LES) are widely adopted in engineering and planetary atmospheric models. The key idea is to take the evolution only down to a specified filtering scale, with the influence of smaller, unresolved subfilter motions taken into account through explicit modeling \citep{Smagorinsky1963,deardorff1970numerical,Germano1992,meneveau2000scale,sagaut2005large}. Because the equations remain unclosed, due to the nonlinear Navier–Stokes operator coupling all scales, empirical models grounded in turbulence theory are employed. Many of these models account for the influence of the small-scale motions through the Reynolds stress tensor $\tau_{ij}$, which captures the action of unresolved scales on the resolved ones. Typically, $\tau_{ij}$ is expressed as a function of the large-scale velocity gradients, consistent with the symmetries of the flow, in order to provide a deterministic model closure for turbulence \citep{meneveau2000scale}. The models by \citet{Smagorinsky1963} and \citet{Clark1979} are some of the most well-known ones. Being quasi-empirical, these models require extensive testing and validation through experiments and simulations \citep{liu1994properties,piomelli1991subgrid,domaradzki1993analysis,hartel1994subgrid}. 
These models have been found to reasonably reproduce simple statistics, however more severe tests reveal often significant discrepancies between modeled and measured stress components~\citep{meneveau2000scale,moser2021statistical}. In particular, looking at the full turbulent cascade process, the eddy-viscosity approximation may lead to oversimplify the picture with large errors on fluctuations. In previous works~\citep{Borue1998,tsinober2009informal,alexakis2018cascades,Alexakis2020},  the local energy flux transferred from unfiltered to filtered scales was investigated. The mean was found to be consistent with the predictions of the simple Smagorinsky model, however significant deviations were observed. 
%It is clear that to obtain a more complete picture of the entire cascade flux is a desirable property of LES models, since of practical importance in some cases.
It is possible that, 
%Such fluctuations are intrinsic to turbulent flows, as even small noise in the initial conditions can lead to significant deviations in finite time \citep{bandak2024spontaneous}. 
since these deviations originate by the small, unresolved scales, their impact on the resolved scales could be captured using stochastic modeling. While stochastic subgrid models have been proposed in previous studies \citep{Carati2007, mason1992stochastic, Marstorp2007}, the statistical properties of the associated noise terms were typically prescribed in an \emph{ad hoc} fashion, and notably are not meant to capture extreme events. 
The need for an accurate stochastic description of the small scale effects is emphasized by recent results proposing that the Navier–Stokes equations can develop randomness at large scales in a finite time~\citep{thalabard2020butterfly,eyink2020renormalization,bandak2024spontaneous}. 
In that view, the development of stochastic models is a necessity for LES models.
%\SC{Perhaps this last section may be expanded}\\
In the present work, we aim to quantify the fluctuations around the mean subgrid-scale energy flux and compare these with the predictions of classical models. To this end, we employ data from direct numerical simulations of the Navier–Stokes equations and follow two complementary approaches: (i) the analysis of several conditional means, as suggested in previous works~\citep{buaria2020self,Johnson2024}; notably we shall analyse probabilities conditioned on QR, which have not been investigated so far. (ii) The joint probability density function (PDF) of model predictions and ground truth, as in \citep{Alexakis2020}, to examine the distribution of subgrid-scale fluxes conditioned on a given model. The ultimate goal is to guide and constrain the development of improved stochastic subgrid-scale models for turbulent flow simulations.

% ----- ===== ----- ===== ----- ===== ----- ===== ----- ===== ----- ===== ----- ===== ----- ===== |

%%%%%%%%%%%%%%%%%%%%%%%%%%%%%%%%%%%%%%%%%%%%%%%%%%%%%%%%%%%%%%%%
\section{Theoretic background}                                %%
\label{sec:tbackground}                                       %%
%%%%%%%%%%%%%%%%%%%%%%%%%%%%%%%%%%%%%%%%%%%%%%%%%%%%%%%%%%%%%%%%

The stage is set by the incompressible Navier–Stokes equations, governing the evolution of a divergence-free velocity field,
%\begin{equation}
%\big[\partial_t-\nu\partial_{jj}\big]u_i+\big[\delta_{ki}-\partial_{ki}\nabla^{-2}\big]\partial_j\left(u_j u_k \right)=0.
%\label{eq:NS}
%\end{equation}
\begin{equation}
\big[\partial_t-\nu\partial_{jj}\big]u_i+
\partial_j\left(u_j u_i \right)-\partial_i p=f_i~,
\label{eq:NS}
\end{equation}
here, $\nu$ is the kinematic viscosity, $p$ is the pressure computed trough incompressibility. An external forcing term is added on the right-hand side, acting at the largest scales of the system. This forcing is statistically homogeneous and injects energy into the flow at a mean rate $\epsilon$. The domain is a triply periodic cubic box of size $2\pi L$, which ensures the absence of boundary-layer effects. We focus on the high Reynolds number regime, with $Re \equiv \epsilon^{1/3} L^{4/3}/\nu$, so that a wide separation of scales emerges between the forcing scale $L$ and the viscous dissipation scale $\eta \equiv \nu^{3/4} \epsilon^{-1/4}$.

\subsection{Filtering approach}
A typical starting point for turbulence modeling is to separate the components describing variations at large and small scales, although their dynamics remain irrevocably coupled because of the nonlinear term. Following \citet{Germano1992} we work with the coarse-graining at scale $l$ of $\vel(\pos,t)$, solution for 3D incompressible Navier–Stokes problem (\ref{eq:NS}), obtained suppressing high-frequency fluctuations through local averaging:
\begin{equation}
    \fil(\pos,t)=\int G^l(\varpos)\vel(\pos-\varpos,t)d\varpos
\end{equation}
where $G^l(\varpos)=l^{-3}G(\varpos/l)$ is a localized smooth filtering function such that $\int G(\varpos)d\varpos=1$. In particular, we use the Gaussian filter, given in Fourier by
\begin{equation}
    \widehat{G}(\scl)=\text{exp}\left[-\frac{k^2}{2}\right].
\end{equation}
The dynamics of the resulting large-scale field is not closed because of the non-linearity:
\begin{equation}
\label{fns}
    \big[\p_t-\nu\p_{jj}\big]\overline{u_i}^l+\big[\delta_{ki}-\p_{ki}\nabla^{-2}\big]\p_j\big(\overline{u_j}^l\overline{u_k}^l\big)=\overline{f_i}^l-\p_j\tau^l(u_i,u_j)
\end{equation}
On the left-hand side, the standard Navier–Stokes operator acts on the coarse-grained field, while the right-hand side depends on the velocity field at all scales through the subgrid-scale tensor
\begin{equation}
\label{sgs}
    \tau^l(u_i,u_j)=\overline{u_i u_j}^l-\overline{u_i}^l\overline{u_j}^l=\vcentcolon\tau^l_{ij},
\end{equation}
the stress exerted on scales larger than $l$ by the chaotic fluctuations at smaller scales. This tensor depends on both the filtered velocity field and the underlying small-scale fluctuations, and it determines the Galilean-invariant local energy transfer from the large-scale to the small-scale component:
\begin{equation}
\label{lef}
    \Pi^l(\pos,t)=-\tau^l_{ij}\p_j \overline{u_i}^l.
\end{equation}
We note that the value of local energy flux here depends on the choice of filter and
technical differences between filters have been noted \cite{Alexakis2020}.
An alternative (filter independent) definition of the energy flux within the structure function framework has been explored \citep{Yao2024}.   
Here we will refrain examining further filters or alternative definitions and 
we will limit ourselves to the Gaussian filter.
A model of turbulence is the closure of equation (\ref{fns}) by expressing the unknown right-hand side in terms of the resolved large-scale field.
Traditionally, the subgrid-scale stress tensor $\tau$ is modeled as a function of the first-order gradients of the filtered velocity. For example, the model by \citet{Clark1979}
\begin{equation}
\label{clark}
    \tau^l_{ij}\sim l^2\p_k\overline{u}_i^l\p_k\overline{u}_j^l~,
\end{equation}
which 
%is a local closure and
may be obtained as a first term of an expansion of (\ref{sgs}) in $l$~\citep{Borue1998,Eyink2006,Johnson2021}. 
Another example is the traditional eddy viscosity model by \citet{Smagorinsky1963}
\begin{equation}
\label{smag.}
    \tau^l_{ij}-\frac{\delta_{ij}}{3}\Tr[\tau^l]\sim-l^2\sqrt{\overline{S}^l_{kh}\overline{S}^l_{hk}}\overline{S}^l_{ij}
\end{equation}
where $S$ is the symmetric part of the velocity gradient tensor, the strain-rate. 
Such models lead to a deterministic evolution of the filtered field. 
%Nevertheless, a key point is that, since the evolution of the small-scale velocity field is not known, and given that the Navier–Stokes equations can develop randomness in a finite time, even from infinitesimally small noisy perturbations \citep{thalabard2020butterfly,eyink2020renormalization,bandak2024spontaneous}, the dynamics of the large-scale field is expected to exhibit an intrinsic stochastic component. Previous studies \citep{Carati2007,Marstorp2007} have attempted to capture this effect by modeling the subgrid-scale stress tensor $\tau$ as a stochastic process.

% ----- ===== ----- ===== ----- ===== ----- ===== ----- ===== ----- ===== ----- ===== ----- ===== |

\subsection{Velocity gradient invariants}
%\ALEX{The basis of most turbulence models is to express $\tau$ through  the velocity gradient tensor $\nabla\vel=\partial_i u_j=s_{ij}-\omega_{ij}$ where $s_{ij}$ is the symmetric and $\omega_{ij}$ the antisymmetric parts of $\nabla\vel$ that isolate the contributions of strain and rotation respectively. }
The tensor 
%\nabla\vel$ 
$\partial_i u_j=S_{ij}+\Omega_{ij}$
(where $S_{ij}$ is the symmetric and $\Omega_{ij}$ the antisymmetric parts of $\nabla\vel$)
encapsulates the spatial variations of the velocity field around a given point and offers key insights into local flow topology, intermittency, and the mechanisms driving the energy cascade.
%
%Following the modeling approach~\citep{Meneveau2011,Johnson2024}, our analysis is based on the velocity gradient tensor $\nabla\vel=A_{ij}=\partial_i u_j=s_{ij}-\omega_{ij}$, which encapsulates the spatial variations of the velocity field around a given point and offers key insights into local flow topology, intermittency, and the mechanisms driving the energy cascade. 
%The decomposition into symmetric ($s_{ij}$) and antisymmetric ($\omega_{ij}$) parts isolates the contributions of strain and rotation. 
%Moreover, 
For isotropic turbulence, $\tau$ should depend only on the properties of $\nabla\vel$ that are invariant under rotations.
In total the gradient tensor is fully characterized by 5 scalar invariant quantities, that are independent of the orientation of the coordinate system. % ~\citep{Meneveau2011,Johnson2024}.
This reflects the fact that the tensor $\nabla\vel$ possesses five effective degrees of freedom, obtained by starting from its nine components and accounting for the incompressibility constraint and the assumption of isotropy \citep{Meneveau2011,Johnson2024}.
To provide a synthetic description of the local fluid configurations without referring directly to the tensor components, 
all 5 invariants are in principle needed. 
However, here we limit ourselves to the two
invariants characterizing the spectrum of $\nabla\vel$.
%Four invariants may be found as coefficients the characteristic equation of an order $3$ traceless real matrix $M$ is the depressed cubic $\lambda^3+Q_M\lambda+R_M=0$, where the coefficients are given by
%Two standard invariants for 
In general, for a traceless real matrix $M$ we will denote
\begin{equation}
	Q_M=-\frac{1}{2}\Tr(M^2),\quad R_M=-\frac{1}{3}\Tr(M^3).
\end{equation}
From now on, we simply refer to $Q$ and $R$ as invariants of the velocity gradient $\nabla\vel$~\citep{Chong1990}:
\begin{equation}
    Q=Q_{{\nabla\vel}}=-\frac{1}{2} \partial_i u_j \partial_i u_j, \quad
    R=R_{{\nabla\vel}}=-\frac{1}{3} \partial_i u_j \partial_j u_k \partial_k u_i.
    %=-1/3S_{ij}S_{jk}S_{ki}-1/4\Omega_{ik}S_{kj}\Omega_{ji}.
\end{equation}
Combining the multiscale description, the invariants of velocity gradient and strain-rate are sufficient to express the energy fluxes predicted by the introduced models (\ref{clark}) and (\ref{smag.}):
\begin{equation}
\label{modqr}\Pi^l_{Clark}=l^2\left[4R_{\overline{S}^l}-R_{\overline{\nabla\vel}^l}\right],\quad \Pi^l_{Smag.}=l^2\left[-Q_{\overline{S}^l}\right]^{3/2}.
\end{equation}
A striking difference between the models is that Smagorinsky's prediction, unlike Clark's, is non-negative.  It is important to note that knowledge of the two invariants of the sole velocity gradient tensor does not uniquely determine those of its symmetric part $S$, and therefore model predictions (\ref{modqr}) are not uniquely defined in the $(Q,R)$ space. %, unlike for the restricted Euler.  
Yet, to link the energy flux to the invariants $Q,R$ has proven useful to get insights on the dynamical process~\citep{Johnson2024}, and also for the construction of LES models~\citep{Borue1998}.

% ----- ===== ----- ===== ----- ===== ----- ===== ----- ===== ----- ===== ----- ===== ----- ===== |

%%%%%%%%%%%%%%%%%%%%%%%%%%%%%%%%%%%%%%%%%%%%%%%%%%%%%%%%%%%%%%%%
\section{Results}                                             %%
\label{sec:results}                                           %%
%%%%%%%%%%%%%%%%%%%%%%%%%%%%%%%%%%%%%%%%%%%%%%%%%%%%%%%%%%%%%%%%
\begin{figure}
  \centerline{\includegraphics{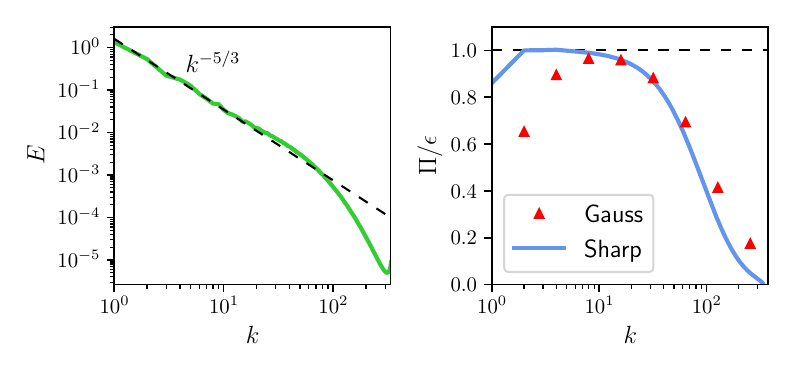}}
  \caption{Kolmogorov energy spectrum (left) and fraction of the mean energy transfer across scales relative to the mean energy injection rate (right).}
\label{fig:1}
\end{figure}
%\subsection{Numerical setup} 
The present study relies on a numerical solution of Navier–Stokes equations (\ref{eq:NS}). 
The pseudo-spectral code GHOST \citep{Mininni2011} was employed, using a fourth-order Runge–Kutta scheme and adapting the time step according to the Courant–Friedrichs–Lewy condition. The code considers a periodic cubic box of size $2\pi$ with resolution $1024$ in each direction. Viscosity was set to $\nu=5\cdot 10^{-4}$. In order to fix the mean energy injection rate $\epsilon$, we imposed the forcing \citep[as, for example, in][]{Alexakis2020}
\begin{equation*}
	\widehat{\boldsymbol{f}}(\boldsymbol{k},t)=\left[\frac{\epsilon}{\scalesum{3ex}_{p\le 2}\vert \widehat{\boldsymbol{u}}(\boldsymbol{p},t)\vert^2}+i\varphi_{\boldsymbol{k}}\right]\widehat{\boldsymbol{u}}(\boldsymbol{k},t).
\end{equation*}
where the phases $\varphi_{\boldsymbol{k}}$ are randomly chosen at initialization and kept fixed in time.
In particular, we put $\epsilon=1$. After a transient, a steady state was reached, statistically consistent with the classical isotropic turbulence. Figure \ref{fig:1} shows, on the left, the energy spectrum, which provides a Kolmogorov-Oboukhov $5/3$ spectrum over almost two decades. On the right, it displays the mean energy transfer $\Pi$ across scale $k$, normalized by the mean energy injection rate. The actual inertial range of scales is identified by the region where the normalized transfer is close to $1$. The blue line represents the spatial and temporal average of (\ref{lef}) using the sharp spectral filter (i.e.,\ the Fourier formulation), while the red triangles correspond to the estimation based on Gaussian filtering. We select the wavenumber $16$ as a representative point for the inertial range.
%\SC{For the sake of completeness, we also show the averaged flux obtained with a sharp filter. As already shown in previous analysis~\citep{Alexakis2020}, the results with the sharp filter present too wide fluctuations allowing a less clear statistical analysis. Moreover, the gaussian filter permits easier analytical developments, so that in the present work we focus only on the gaussian filter.}
% ----- ===== ----- ===== ----- ===== ----- ===== ----- ===== ----- ===== ----- ===== ----- ===== |

\subsection{Clark vs DNS}
\subsubsection{QR diagrams}

To compare \textit{a priori} predictions given by models of turbulence we propose to look at the conditional mean $\langle\Pi\vert Q,R\rangle$. 
This observable has been first used in a recent work~\citep{vela2021entropy}, yet in a different context.
\begin{figure}
\centerline{\includegraphics{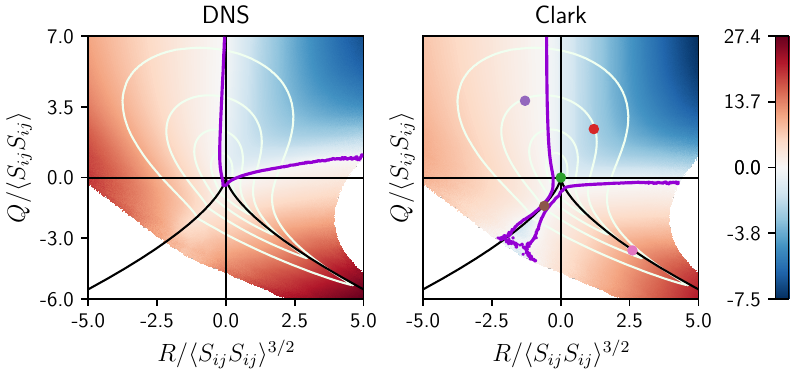}}
  \caption{Color scale represents the mean local energy flux conditioned on $(Q,R)$ configurations at filtering scale $q=16$. Positive fluxes (forward cascade) are shown in red, and negative fluxes (backscatter) in blue. The left panel displays DNS results, while the right panel shows the Clark model prediction. White contours indicate isolines of the probability density in the $(Q,R)$ plane at levels 10$^{-4}$, 10$^{-5}$, 10$^{-6}$ and 10$^{-7}$, moving outward from the origin. The purple curve denotes the isoline of zero flux. Colored dots in the right panel indicate the $(Q,R)$ locations used for the conditional analysis in figure \ref{fig:3}.}
\label{fig:2}
\end{figure}
In figure \ref{fig:2} we consider $Q$ and $R$ normalized by the mean dissipation at the corresponding scale, and showing the mean flux for a given $(Q,R)$ with the color scale: positive values in red indicate a mean energy transfer to smaller scales, while negative values in blue correspond to a net transfer toward larger scales. White contours indicate the isolines of the probability density on the $(Q,R)$ space, corresponding to levels 10$^{-4}$, 10$^{-5}$, 10$^{-6}$, and 10$^{-7}$, starting from the curve closest to the origin, which corresponds to the maximum of the distribution. 
The violet curve marks the isoline where the mean flux vanishes. The left panel presents the results obtained from Eq. (\ref{lef}), whereas the right panel shows the prediction of the Clark model (\ref{clark}), calculated using the complete DNS fields. 
First, it is noteworthy that, in DNS, the backward cascade (backscatter) is  confined to the upper-right quadrant, associated with vortex compression configurations. 
Then, the Clark model successfully captures the qualitative distribution of the mean energy flux across the $(Q,R)$ diagram. However, quantitatively, it underestimates the right value. It is possible to see that a region with small backscatter flow is observable in the lower-left panel, as highlighted by the violet curve.
%\begin{figure}
%  \centerline{\includegraphics[scale=0.75]{nonLoc.pdf}}
%  \caption{Difference in the local energy flux PDFs conditioned on $(Q,R)$  between ground-truth DNS and Clark model predictions. }
%\label{fig:2bis}
%\end{figure}
We can discuss a little more the error made by the Clark model from a physical point of view.
As discussed in previous works~\citep{Eyink2006}, the  turbulent subgrid stress may be written as an infinite series of terms in which the Clark model represents the first term.
In particular, for a gaussian filter as the one used here, it may be written~\citep{Johnson2021}:
\begin{equation}
\tau^l_{ij}\sim l^2\p_k\overline{u}_i^l\p_k\overline{u}_j^l
+\int_0^{\ell^2} d\theta 
\left(\overline{\overline{\nabla_i \vel_k}^{\sqrt{\theta}} 
                \overline{\nabla_j \vel_k}^{\sqrt{\theta}}}^{\phi}
-\overline{     \overline{\nabla_i \vel_k}^{\sqrt{\theta}}}^{\phi} ~
\overline{      \overline{\nabla_j \vel_k}^{\sqrt{\theta}}}^{\phi}
\right)~,
\label{eq:john}
\end{equation}
where $\phi=\sqrt{\ell^2-\theta}$.
The first term is scale-local and corresponds to the Clark model, which therefore neglects the higher-order nonlocal terms involving scales smaller than the filter one. The difference between the Clark model and DNS in the energy flux is related to this term.
% In Fig. \ref{fig:2bis}, we show the  difference in the mean flux $\Pi_{DNS}-\Pi_{Clark}$ conditioned for $QR$.
We emphasize that the discussion is made at the filter scale $q=16$, that is in the inertial range.
We can see that the local-in-scale model globally underestimates the mean.
%, except in the upper right panel where vortex compression is dominant.
\begin{figure}
  \centerline{\includegraphics{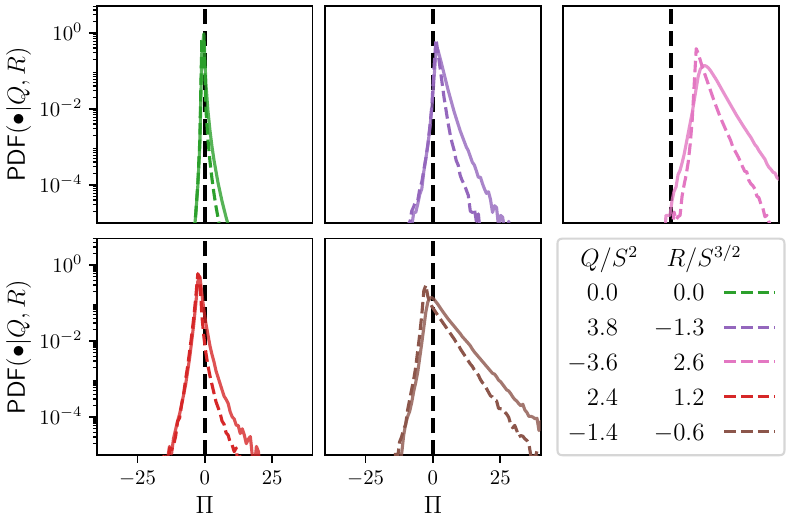}}
  \caption{Comparison of local energy flux PDFs conditioned on $(Q,R)$ configurations, between ground-truth DNS (solid lines) and Clark model predictions (dashed lines). The corresponding $(Q,R)$ locations are indicated in figure \ref{fig:2}.}
\label{fig:3}
\end{figure}
In figure \ref{fig:3}, we move beyond the analysis of the mean and examine the distribution of the Clark prediction versus the ground-truth energy flux conditioned on five representative points in the $(Q,R)$-space: the origin and one point in each of the four quadrants, highlighted in figure \ref{fig:2}. 
This assessment is severe, but it permits to fully recognize the capability to reproduce the cascade process by a model.
This type of conditioning is particularly interesting since it corresponds to imposing a constraint on the large-scale flow topology.
The solid lines show the conditioned PDF for the measured energy flux directly from the DNS, while the dashed
lines give the conditioned PDFs for the predicted energy flux from the Clark model. We remind the $Q,R$ do not uniquely 
determine the energy flux in the Clark model and that for the same $Q,R$ a distribution of energy fluxes is observed.
Globally speaking, the Clark model makes a decent job overall. The essential features of the flux are captured, and in particular we observe a good agreement in the negative/backscatter tail.
Nonetheless, the skewness is not correct and the positive tail is systematically underestimated. 
The shape of the pdf appears similar, but a more careful look at the right tails shows that the scaling is also different.
Overall, stronger fluctuations are observed in the DNS than the Clark model predictions.
Considering the decomposition in scales (\ref{eq:john}), the nonlocal terms appear important in the positive tails of the probability of the flux.

% ----- ===== ----- ===== ----- ===== ----- ===== ----- ===== ----- ===== ----- ===== ----- ===== |

\subsubsection{Joint PDFs}
%We proceed by extending the analysis beyond the mean. 
To become more precise we introduce the joint probability density of local energy flux (\ref{lef}) and Clark model prediction (\ref{clark}). This joint PDF serves as the appropriate observable to assess the fluctuations around the model: for a fixed value of the Clark prediction, the complete flux exhibits a distribution of values, reflecting the influence of the chaotic small-scale dynamics. Our primary objective is to investigate the significance of these fluctuations.
%, with particular focus on their non-Gaussian nature. 
\begin{figure}
  \centerline{\includegraphics{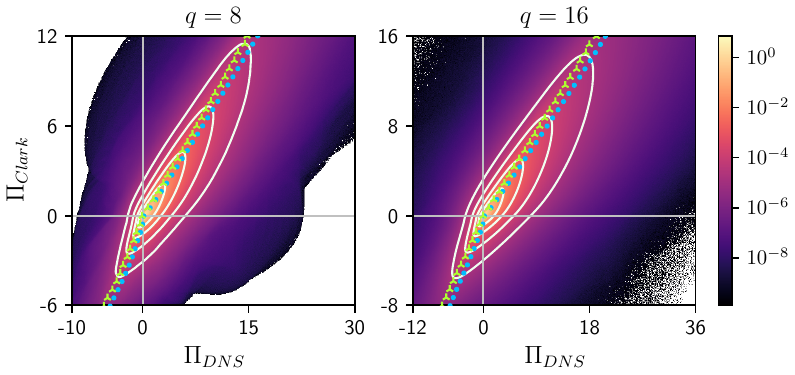}}
  \caption{Color scale represents the joint probability density functions of \ref{lef} and \ref{clark}. White contours indicate probability isolines at levels 10$^{-4}$, 10$^{-5}$, 10$^{-6}$, and 10$^{-7}$, progressing outward from the origin. Green upward tripod markers denote the peaks of the conditional PDFs obtained by fixing the model prediction (horizontal cuts), while blue circles represent the corresponding mean values.}
\label{fig:4}
\end{figure}
In figure \ref{fig:4} we present this joint PDF. The white contours have the same meaning as in the previous plot. Green markers indicate the maxima of the conditional PDFs for fixed model values (horizontal cuts), while blue circles represent the corresponding mean values. The discrepancies between the mode and the mean provide a first indication of non-Gaussian behavior. Yet, the difference is small.
Furthermore, to a good approximation the maxima stay on a linear curve,  although the slope is slightly different in the negative and positive range.
These evidences corroborate the previous conclusion that the Clark model performs well with respect to the most probable events. In the inertial range, the linear behavior with slope of order one shows that the contribution of the nonlocal terms are negligible with respect to the mean flux~\citep{Johnson2021}.  
\begin{figure}
  \centerline{\includegraphics{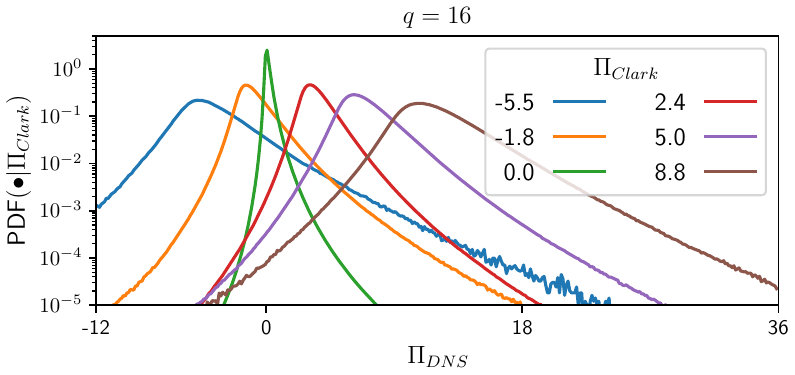}}
  \caption{PDFs of the \ref{lef} energy flux, conditioned to values for the Clark estimation \ref{clark}.}
\label{fig:5}
\end{figure}
Examples of selected conditional distributions, given the model prediction (i.e.,\ the resolved field), are presented in figure \ref{fig:5}. These correspond to horizontal cuts in figure \ref{fig:4}. 
Large fluctuations with a variance of the order of the mean are observed. Then, the pdfs are characterised by pronounced non-Gaussian fat tails for each value of the flux . These heavy-tailed statistics indicate the occurrence of rare but intense deviations from the average behavior, highlighting the intermittent and strongly non-linear nature of the system. The observed behavior suggests that the tails could follow a distribution intermediate between exponential and power-law decay.  Yet, it appears that there is not an universal shape for the pdf, and we have not been able to rescale toward a sole master curve 
%\ALEX{possibly due to the limited $Re$ examined here.}
%\SC{Flavio: could you make a collapse of the curve by rescaling to see the differences?}
To quantify these observations, we compute the first four cumulants of the conditional PDFs, varying the value of the model. These are mean and variance,
\begin{equation}
    \mu=\int\Pi\,d\mathbb{P}[\Pi\vert\Pi_{model}],\quad
    \sigma^2=\int(\Pi-\mu)^2\,d\mathbb{P}[\Pi\vert\Pi_{model}],
\end{equation}
skewness and flatness,
\begin{equation}
    \gamma_1=\frac{1}{\sigma^3}\int(\Pi-\mu)^3\,d\mathbb{P}[\Pi\vert\Pi_{model}],\quad
    \gamma_2=\frac{1}{\sigma^4}\int(\Pi-\mu)^4\,d\mathbb{P}[\Pi\vert\Pi_{model}],
\end{equation}
where $\Pi$ denotes the DNS value (\ref{lef}).
\begin{figure}
  \centerline{\includegraphics{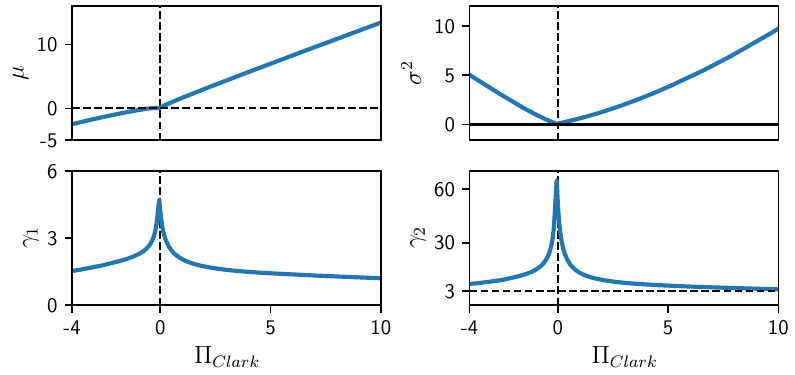}}
  \caption{Statistical cumulants (mean, variance, skewness, and flatness) of the DNS density, computed conditionally on a value of the Clark flux.}
\label{fig:6}
\end{figure}
The conditional mean suggests that a linear fit is appropriate, indicating that the expression (\ref{clark}) could provide a good approximation when adjusted by a multiplicative constant. It is also worth noting that, on average, the DNS flux shares the same sign as the Clark model prediction. However, the slopes differ between the positive and negative branches. Continuing the analysis, and interpreting the variance as a measure of statistical uncertainty, we observe that extreme events (i.e.,\ those with large absolute values) are associated with greater variability. Interestingly, the ratio $\sigma/\mu$ appears to decrease as one moves away from zero, pointing toward the deterministic approximation in the large $\Pi_{Clark}$ limit. The skewness, however, clearly indicates that the distribution remains not symmetric. Gaussian flatness is recovered for extreme Clark model predictions, whereas fat-tailed behavior dominates near the most probable events.

% ----- ===== ----- ===== ----- ===== ----- ===== ----- ===== ----- ===== ----- ===== ----- ===== |

\subsection{Smagorinsky vs DNS}
\begin{figure}
  \centerline{\includegraphics{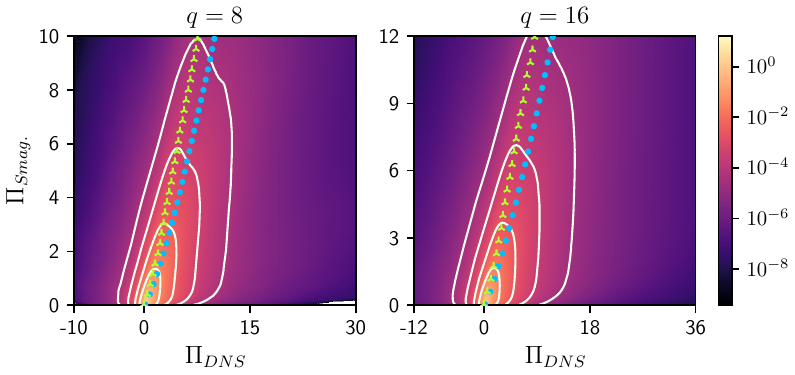}}
  \caption{Color scale represents the joint probability density functions of \ref{lef} and \ref{smag.}. White contours indicate probability isolines at levels 10$^{-4}$, 10$^{-5}$, 10$^{-6}$, and 10$^{-7}$, progressing outward from the origin. Green upward tripod markers denote the peaks of the conditional PDFs obtained by fixing the model prediction (horizontal cuts), while blue circles represent the corresponding mean values.}
\label{fig:7}
\end{figure}

\begin{figure}
  \centerline{\includegraphics{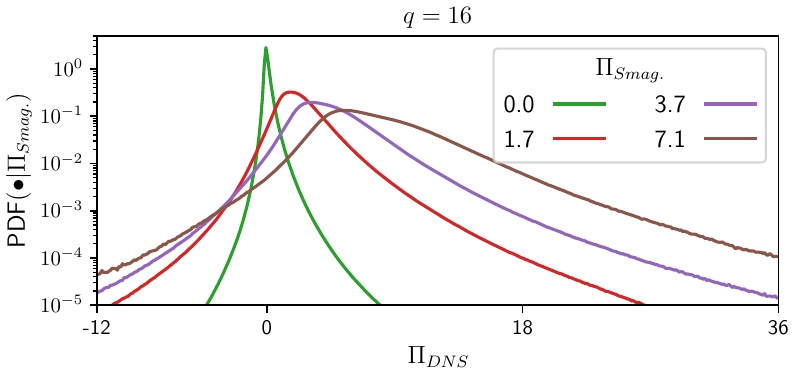}}
  \caption{PDFs of the \ref{lef} energy flux, conditioned to Smagorinsky estimations \ref{smag.}.}
\label{fig:8}
\end{figure}

For completeness we perform a parallel analysis for the classical eddy-viscosity model of Smagorinsky, described by expression (\ref{smag.}). Unlike the Clark model, the Smagorinsky formulation predicts only non-negative energy flux, thereby completely failing to account for the backscatter. In figure \ref{fig:7}, we present the joint PDF of ground-truth and model prediction: a multiplicative constant can be fitted to align either the maximum probability or the mean value.
As in figure \ref{fig:5}, the figure \ref{fig:8} displays the conditional probability density functions of the DNS energy flux for fixed values of the Smagorinsky output. These distributions exhibit prominent heavy tails, characteristic of rare but intense events. A fit of the conditional distributions reveals that the tails decay more slowly than exponentially, supporting the idea that rare events dominate the flux statistics and cannot be neglected in closure strategies.
To further quantify these features, we look at the first four statistical cumulants of the conditional distributions, shown in figure \ref{fig:8}.
 \begin{figure}
  \centerline{\includegraphics{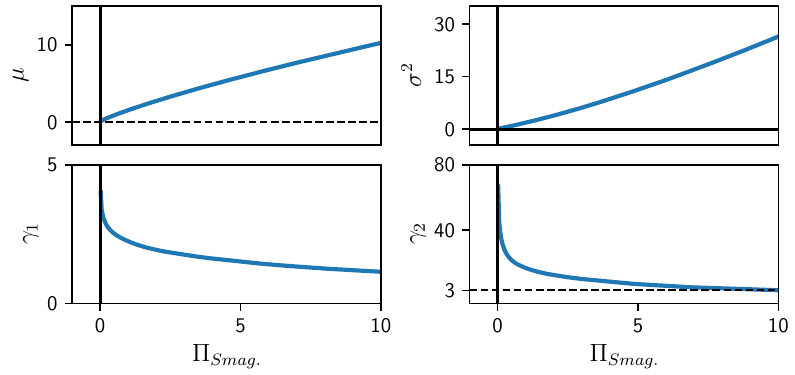}}
  \caption{Statistical cumulants (mean, variance, skewness, and flatness) of the DNS density, computed conditionally on a threshold in the Smagorinsky model response.}
\label{fig:9}
\end{figure}
As with the Clark model, the conditional mean has a linear trend versus the model prediction. Yet the variance increases with increasing model flux much more for the Smagorinsky case, indicating greater uncertainty for more intense events. However, in that regime, the relative uncertainty $\sigma/\mu$ approaches zero. The skewness exhibits a pronounced positive trend, highlighting the strong asymmetry of the distributions. As for the flatness, values close to the Gaussian reference (i.e.,\ 3) are approached at large predicted values, while extreme flatness, indicative of intermittent fat-tailed fluctuations, dominates regions with moderate model output. These findings confirm the limited predictive capability of the Smagorinsky model in capturing the detailed structure of subgrid-scale energy transfer, and further illustrate the breakdown of Gaussian assumptions near the Smagorinsky prediction.

% ----- ===== ----- ===== ----- ===== ----- ===== ----- ===== ----- ===== ----- ===== ----- ===== |

\section{Discussion}
\label{sec:discussion}

We have investigated different conditional statistics of the filtered 
turbulent velocity field in 3D. The analysis of the mean energy flux in the QR diagram shows unambiguously that the backscatter inverse cascade is correlated solely to the vortex compression, while positive flux is due to all quadrants. Results highlight the role of the nonlocal neglected terms, which are found to be responsible of an increase of the positive extreme  events  of the energy flux.
With respect to the LES framework, we have found that the nonlinear Clark model reproduces satisfactorily  the mean energy flux and provides an approximate reconstruction of the 
fluctuations around the mean, but it remains insufficient to fully capture the complex statistical structure of the energy flux in turbulent flows. As expected, the Smagorinsky model is found to give only a very crude description of the cascade phenomenon.
While the Clark model can be considered a good first order approximation,  we have highlighted the presence of heavy-tailed distributions and marked asymmetries, hallmarks of the chaotic and intermittent nature of 
the modeled unresolved scales. 
%The results presented in this study confirm that deterministic subgrid-scale models, such as those of Clark and Smagorinsky, are insufficient to fully capture the complex statistical structure of the energy flux in turbulent flows. Through the analysis of the conditional dependence of the DNS flux on the model predictions, along with the behavior of the associated cumulants, we have highlighted the presence of heavy-tailed distributions and marked asymmetries, hallmarks of the chaotic and intermittent nature of \ALEX{the modeled unresolved scales}.
These findings underscore the limitations of classical deterministic closures based solely on filtered quantities and point toward the need for stochastic components to better account for the influence of unresolved scales. 
While deterministic models may reproduce the average trend, as shown by the QR-conditioned analysis, large fluctuations persist around the mean. 
The ratio $\sigma/\mu$ decreases with increasing the flux value, suggesting a transition toward a gaussian limit for the high flux events. 
However, the strong skewness and enhanced flatness observed at moderate model outputs demonstrate that rare events and intermittent bursts remain statistically significant and cannot be neglected. In this context, fluctuations in the subgrid stress tensor can not be modeled by simple Gaussian noise. More general noise than Gaussian needs to be introduced in models than what is done in the past and possibly more complex coupling between the strain tensor and the imposed noise.
%The present results can provide a basis that can constrain future stochastic models}.  
Starting from available stochastic modelling for LES~\citep{gicquel2002velocity}, our results provide a quantitative foundation for the development of stochastic subgrid-scale models that can incorporate both the mean behavior and the statistical variability of energy transfer.
Finally we need to remark that
in this work, we have focused on stationary statistics to construct time-independent PDFs. An important direction for future research is to investigate temporal correlations, with the aim of constructing fully dynamical stochastic processes for the subgrid-scale energy flux.

\backsection[Acknowledgements]{This work received financial support from the CNRS through the MITI interdisciplinary initiatives, under its exploratory research program. SC acknowledges the funding by ANR SPEED  project ANR-20-CE23-0025-01.
It was granted access to the HPC resources of GENCI-TGCC \& GENCI-
CINES (Project No. A0170506421).
}

% ----- ===== ----- ===== ----- ===== ----- ===== ----- ===== ----- ===== ----- ===== ----- ===== |

%\appendix
%\section{}

%\begin{figure}
 % \centerline{\includegraphics{qrpCutErr}}
  %\caption{PDFs of the pointwise signed error made by the Clark model in predicting the energy flux, conditioned on $(Q,R)$ configurations.}
%\label{fig:7}
%\end{figure}

% ----- ===== ----- ===== ----- ===== ----- ===== ----- ===== ----- ===== ----- ===== ----- ===== |

\bibliographystyle{jfm}
\bibliography{jfm}

\begin{thebibliography}{36}
\expandafter\ifx\csname natexlab\endcsname\relax\def\natexlab#1{#1}\fi
\def\au#1{#1} \def\ed#1{#1} \def\yr#1{#1}\def\at#1{#1}\def\jt#1{\textit{#1}}
  \def\bt#1{#1}\def\bvol#1{\textbf{#1}} \def\vol#1{#1} \def\pg#1{#1}
  \def\publ#1{#1}\def\arxiv#1{#1}\def\org#1{#1}\def\st#1{\textit{#1}}

\bibitem[Alexakis \& Biferale(2018)]{alexakis2018cascades}
{\sc \au{Alexakis, A.} \& \au{Biferale, L.}} \yr{2018}  \at{Cascades and
  transitions in turbulent flows}.  \jt{Physics Reports}  \bvol{767},
  \pg{1--101}.

\bibitem[Alexakis \& Chibbaro(2020)]{Alexakis2020}
{\sc \au{Alexakis, A.} \& \au{Chibbaro, S.}} \yr{2020}  \at{On the local energy
  flux of turbulent flows}.  \jt{Physical Review Fluids}  \bvol{5}~(9).

\bibitem[Bandak {\em et~al.\/}(2024)Bandak, Mailybaev, Eyink \&
  Goldenfeld]{bandak2024spontaneous}
{\sc \au{Bandak, D.}, \au{Mailybaev, A.~A.}, \au{Eyink, G.~L.} \&
  \au{Goldenfeld, N.}} \yr{2024}  \at{Spontaneous stochasticity amplifies even
  thermal noise to the largest scales of turbulence in a few eddy turnover
  times}.  \jt{Physical review letters}  \bvol{132}~(10),  \pg{104002}.

\bibitem[Biskamp(2003)]{biskamp2003magnetohydrodynamic}
{\sc \au{Biskamp, D.}} \yr{2003} {\em Magnetohydrodynamic turbulence\/}.
  \publ{Cambridge University Press}.

\bibitem[Borue \& Orszag(1998)]{Borue1998}
{\sc \au{Borue, V.} \& \au{Orszag, S.~A.}} \yr{1998}  \at{Local energy flux and
  subgrid-scale statistics in three-dimensional turbulence}.  \jt{Journal of
  Fluid Mechanics}  \bvol{366},  \pg{1--31}.

\bibitem[Buaria {\em et~al.\/}(2020)Buaria, Pumir \&
  Bodenschatz]{buaria2020self}
{\sc \au{Buaria, D.}, \au{Pumir, A.} \& \au{Bodenschatz, E.}} \yr{2020}
  \at{Self-attenuation of extreme events in navier--stokes turbulence}.
  \jt{Nature communications}  \bvol{11}~(1),  \pg{5852}.

\bibitem[Carati {\em et~al.\/}(2007)Carati, Winckelmans \&
  Jeanmart]{Carati2007}
{\sc \au{Carati, D.}, \au{Winckelmans, G.} \& \au{Jeanmart, H.}} \yr{2007}
  \at{A stochastic subgrid model with application to turbulent flow and scalar
  transport}.  \jt{Physics of Fluids}  \bvol{19}~(3),  \pg{035107}.

\bibitem[Chong {\em et~al.\/}(1990)Chong, Perry \& Cantwell]{Chong1990}
{\sc \au{Chong, M.~S.}, \au{Perry, A.~E.} \& \au{Cantwell, B.~J.}} \yr{1990}
  \at{A general classification of three-dimensional flow fields}.  \jt{Physics
  of Fluids A: Fluid Dynamics}  \bvol{2}~(5),  \pg{765--777}.

\bibitem[Clark {\em et~al.\/}(1979)Clark, Ferziger \& Reynolds]{Clark1979}
{\sc \au{Clark, R.~A.}, \au{Ferziger, J.~H.} \& \au{Reynolds, W.~C.}} \yr{1979}
   \at{Evaluation of subgrid-scale models using an accurately simulated
  turbulent flow}.  \jt{Journal of Fluid Mechanics}  \bvol{91}~(1),
  \pg{1--16}.

\bibitem[Deardorff(1970)]{deardorff1970numerical}
{\sc \au{Deardorff, J.~W.}} \yr{1970}  \at{A numerical study of
  three-dimensional turbulent channel flow at large reynolds numbers}.
  \jt{Journal of Fluid Mechanics}  \bvol{41}~(2),  \pg{453--480}.

\bibitem[Domaradzki {\em et~al.\/}(1993)Domaradzki, Liu \&
  Brachet]{domaradzki1993analysis}
{\sc \au{Domaradzki, J.~A.}, \au{Liu, W.} \& \au{Brachet, M.~E.}} \yr{1993}
  \at{An analysis of subgrid-scale interactions in numerically simulated
  isotropic turbulence}.  \jt{Physics of Fluids A: Fluid Dynamics}
  \bvol{5}~(7),  \pg{1747--1759}.

\bibitem[Eyink(2006)]{Eyink2006}
{\sc \au{Eyink, G.~L.}} \yr{2006}  \at{Multi-scale gradient expansion of the
  turbulent stress tensor}.  \jt{Journal of Fluid Mechanics}  \bvol{549},
  \pg{159--190}.

\bibitem[Eyink \& Bandak(2020)]{eyink2020renormalization}
{\sc \au{Eyink, G.~L.} \& \au{Bandak, D.}} \yr{2020}  \at{Renormalization group
  approach to spontaneous stochasticity}.  \jt{Physical Review Research}
  \bvol{2}~(4),  \pg{043161}.

\bibitem[Germano(1992)]{Germano1992}
{\sc \au{Germano, M.}} \yr{1992}  \at{Turbulence: the filtering approach}.
  \jt{Journal of Fluid Mechanics}  \bvol{238},  \pg{325--336}.

\bibitem[Gicquel {\em et~al.\/}(2002)Gicquel, Givi, Jaberi \&
  Pope]{gicquel2002velocity}
{\sc \au{Gicquel, L. Y.~M.}, \au{Givi, P.}, \au{Jaberi, F.~A.} \& \au{Pope,
  S.~B.}} \yr{2002}  \at{Velocity filtered density function for large eddy
  simulation of turbulent flows}.  \jt{Physics of Fluids}  \bvol{14}~(3),
  \pg{1196--1213}.

\bibitem[Gill(2016)]{gill2016atmosphere}
{\sc \au{Gill, A.~E.}} \yr{2016} {\em Atmosphere—ocean dynamics\/}.
  \publ{Elsevier}.

\bibitem[H{\"a}rtel {\em et~al.\/}(1994)H{\"a}rtel, Kleiser, Unger \&
  Friedrich]{hartel1994subgrid}
{\sc \au{H{\"a}rtel, C.}, \au{Kleiser, L.}, \au{Unger, F.} \& \au{Friedrich,
  R.}} \yr{1994}  \at{Subgrid-scale energy transfer in the near-wall region of
  turbulent flows}.  \jt{Physics of Fluids}  \bvol{6}~(9),  \pg{3130--3143}.

\bibitem[Johnson(2021)]{Johnson2021}
{\sc \au{Johnson, P.~L.}} \yr{2021}  \at{On the role of vorticity stretching
  and strain self-amplification in the turbulence energy cascade}.  \jt{Journal
  of Fluid Mechanics}  \bvol{922},  \pg{A3}.

\bibitem[Johnson \& Wilczek(2024)]{Johnson2024}
{\sc \au{Johnson, P.~L.} \& \au{Wilczek, M.}} \yr{2024}  \at{Multiscale
  velocity gradients in turbulence}.  \jt{Annual Review of Fluid Mechanics}
  \bvol{56},  \pg{463--490}.

\bibitem[Liu {\em et~al.\/}(1994)Liu, Meneveau \& Katz]{liu1994properties}
{\sc \au{Liu, S.}, \au{Meneveau, C.} \& \au{Katz, J.}} \yr{1994}  \at{On the
  properties of similarity subgrid-scale models as deduced from measurements in
  a turbulent jet}.  \jt{Journal of Fluid Mechanics}  \bvol{275},
  \pg{83--119}.

\bibitem[Marstorp {\em et~al.\/}(2007)Marstorp, Brethouwer \&
  Johansson]{Marstorp2007}
{\sc \au{Marstorp, L.}, \au{Brethouwer, G.} \& \au{Johansson, A.~V.}} \yr{2007}
   \at{A stochastic subgrid model with application to turbulent flow and scalar
  mixing}.  \jt{Physics of fluids}  \bvol{19}~(3).

\bibitem[Mason \& Thomson(1992)]{mason1992stochastic}
{\sc \au{Mason, P.~J.} \& \au{Thomson, D.~J.}} \yr{1992}  \at{Stochastic
  backscatter in large-eddy simulations of boundary layers}.  \jt{Journal of
  Fluid Mechanics}  \bvol{242},  \pg{51--78}.

\bibitem[Meneveau(2011)]{Meneveau2011}
{\sc \au{Meneveau, C.}} \yr{2011}  \at{Lagrangian dynamics and models of the
  velocity gradient tensor in turbulent flows}.  \jt{Annual Review of Fluid
  Mechanics}  \bvol{43},  \pg{219--245}.

\bibitem[Meneveau \& Katz(2000)]{meneveau2000scale}
{\sc \au{Meneveau, C.} \& \au{Katz, J.}} \yr{2000}  \at{Scale-invariance and
  turbulence models for large-eddy simulation}.  \jt{Annual Review of Fluid
  Mechanics}  \bvol{32}~(1),  \pg{1--32}.

\bibitem[Mininni {\em et~al.\/}(2011)Mininni, Rosenberg, Reddy \&
  Pouquet]{Mininni2011}
{\sc \au{Mininni, P.~D.}, \au{Rosenberg, D.}, \au{Reddy, R.} \& \au{Pouquet,
  A.}} \yr{2011}  \at{A hybrid mpi–openmp scheme for scalable parallel
  pseudospectral computations for fluid turbulence}.  \jt{Parallel Computing}
  \bvol{37}~(6–7),  \pg{316--326}.

\bibitem[Moin \& Mahesh(1998)]{moin1998direct}
{\sc \au{Moin, P.} \& \au{Mahesh, K.}} \yr{1998}  \at{Direct numerical
  simulation: a tool in turbulence research}.  \jt{Annual review of fluid
  mechanics}  \bvol{30}~(1),  \pg{539--578}.

\bibitem[Moser {\em et~al.\/}(2021)Moser, Haering \&
  Yalla]{moser2021statistical}
{\sc \au{Moser, R.~D.}, \au{Haering, S.~W.} \& \au{Yalla, G.~R.}} \yr{2021}
  \at{Statistical properties of subgrid-scale turbulence models}.  \jt{Annual
  Review of Fluid Mechanics}  \bvol{53}~(1),  \pg{255--286}.

\bibitem[Piomelli {\em et~al.\/}(1991)Piomelli, Cabot, Moin \&
  Lee]{piomelli1991subgrid}
{\sc \au{Piomelli, U.}, \au{Cabot, W.~H.}, \au{Moin, P.} \& \au{Lee, S.}}
  \yr{1991}  \at{Subgrid-scale backscatter in turbulent and transitional
  flows}.  \jt{Physics of Fluids A: Fluid Dynamics}  \bvol{3}~(7),
  \pg{1766--1771}.

\bibitem[Pope(2001)]{pope2001turbulent}
{\sc \au{Pope, S.~B.}} \yr{2001}  \at{Turbulent flows}.  \jt{Measurement
  Science and Technology}  \bvol{12}~(11),  \pg{2020--2021}.

\bibitem[Sagaut(2005)]{sagaut2005large}
{\sc \au{Sagaut, P.}} \yr{2005} {\em Large eddy simulation for incompressible
  flows: an introduction\/}.  \publ{Springer Science \& Business Media}.

\bibitem[Smagorinsky(1963)]{Smagorinsky1963}
{\sc \au{Smagorinsky, J.}} \yr{1963}  \at{General circulation experiments with
  the primitive equations: I. the basic experiment}.  \jt{Monthly Weather
  Review}  \bvol{91}~(3),  \pg{99--164}.

\bibitem[Thalabard {\em et~al.\/}(2020)Thalabard, Bec \&
  Mailybaev]{thalabard2020butterfly}
{\sc \au{Thalabard, S.}, \au{Bec, J.} \& \au{Mailybaev, A.~A.}} \yr{2020}
  \at{From the butterfly effect to spontaneous stochasticity in singular shear
  flows}.  \jt{Communications Physics}  \bvol{3}~(1),  \pg{122}.

\bibitem[Thorpe(2007)]{thorpe2007introduction}
{\sc \au{Thorpe, S.~A.}} \yr{2007} {\em An introduction to ocean turbulence\/},
  ,  \vol{vol.~10}.  \publ{Cambridge University Press Cambridge}.

\bibitem[Tsinober(2009)]{tsinober2009informal}
{\sc \au{Tsinober, A.}} \yr{2009} {\em An informal conceptual introduction to
  turbulence\/}.  \publ{Springer}.

\bibitem[Vela-Mart{\'\i}n \& Jim{\'e}nez(2021)]{vela2021entropy}
{\sc \au{Vela-Mart{\'\i}n, A.} \& \au{Jim{\'e}nez, J.}} \yr{2021}  \at{Entropy,
  irreversibility and cascades in the inertial range of isotropic turbulence}.
  \jt{Journal of Fluid Mechanics}  \bvol{915},  \pg{A36}.

\bibitem[Yao {\em et~al.\/}(2024)Yao, Schnaubelt, Szalay, Zaki \&
  Meneveau]{Yao2024}
{\sc \au{Yao, H.}, \au{Schnaubelt, M.}, \au{Szalay, A.~S.}, \au{Zaki, T.~A.} \&
  \au{Meneveau, C.}} \yr{2024}  \at{Comparing local energy cascade rates in
  isotropic turbulence using structure-function and filtering formulations}.
  \jt{Journal of Fluid Mechanics}  \bvol{980},  \pg{A42}.

\end{thebibliography}

\end{document}